\PassOptionsToPackage{bookmarks=false}{hyperref}
\documentclass[conference]{IEEEtran}
\usepackage[font=small,labelsep=space]{caption}
\usepackage{verbatim}
\usepackage{graphicx}
\usepackage{multirow}
\usepackage{amssymb}
\usepackage{amsmath}
\usepackage{amsthm}
\usepackage{amsfonts}
\usepackage{paralist}
\usepackage{url}
\usepackage{epstopdf}
\usepackage{float}
\usepackage{color}
\usepackage{cite}
\usepackage{paralist}
\usepackage[noend]{algpseudocode}
\usepackage{algorithmicx}
\usepackage{algorithm}
\usepackage{mathrsfs}
\usepackage{amsmath}
\usepackage{caption}
\usepackage{subfigure}
\usepackage{hyperref}
\hypersetup{hidelinks}
\let\itemize\compactitem
\let\enditemize\endcompactitem
\let\enumerate\compactenum
\let\endenumerate\endcompactenum

\usepackage{algpseudocode}

\makeatletter
\renewcommand{\maketag@@@}[1]{\hbox{\m@th\normalsize\normalfont#1}}
\makeatother

\begin{document}
\title{UAV-Assisted Image Acquisition: 3D UAV Trajectory Design and Camera Control} 
\makeatletter
\newcommand{\linebreakand}{%
  \end{@IEEEauthorhalign}
  \hfill\mbox{}\par
  \mbox{}\hfill\begin{@IEEEauthorhalign}
}
\makeatother

\author{Xiao-Wei~Tang\IEEEauthorrefmark{1}, \emph{Member, IEEE}, Shuowen Zhang\IEEEauthorrefmark{2}, \emph{Member, IEEE}, Changsheng You\IEEEauthorrefmark{3}, \emph{Member, IEEE}, \\
Xin-Lin Huang\IEEEauthorrefmark{1}, \emph{Senior Member, IEEE}, and Rui Zhang\IEEEauthorrefmark{4}, \emph{Fellow, IEEE} \\
\IEEEauthorrefmark{1}Department of Information and Communication Engineering, Tongji University \\
\IEEEauthorrefmark{2}Department of Electronic and Information Engineering, The Hong Kong Polytechnic University \\
\IEEEauthorrefmark{3}Department of Electrical and Electronic Engineering, Southern University of Science and Technology \\
\IEEEauthorrefmark{4}Department of Electrical and Computer Engineering, National University of Singapore \\
%\IEEEauthorrefmark{1}\{xwtang, xlhuang\}@tongji.edu.cn, \IEEEauthorrefmark{2}shuowen.zhang@polyu.edu.hk, \\\IEEEauthorrefmark{3}youcs@sustech.edu.cn, \IEEEauthorrefmark{4}elezhang@nus.edu.sg
Email: \IEEEauthorrefmark{1}\{xwtang, xlhuang\}@tongji.edu.cn, \IEEEauthorrefmark{2}shuowen.zhang@polyu.edu.hk, \\\IEEEauthorrefmark{3}youcs@sustech.edu.cn, \IEEEauthorrefmark{4}elezhang@nus.edu.sg
}
\maketitle

\begin{abstract}
In this paper, we consider a new unmanned aerial vehicle (UAV)-assisted oblique image acquisition system where a UAV is dispatched to take images of multiple ground targets (GTs). To study the three-dimensional (3D) UAV trajectory design for image acquisition, we first propose a novel UAV-assisted oblique photography model, which characterizes the image resolution with respect to the UAV's 3D image-taking location. Then, we formulate a 3D UAV trajectory optimization problem to minimize the UAV's traveling distance subject to the image resolution constraints. The formulated problem is shown to be equivalent to a modified 3D traveling salesman problem with neighbourhoods, which  is NP-hard in general. To tackle this difficult problem, we propose an iterative algorithm to obtain a high-quality suboptimal solution efficiently, by alternately optimizing the UAV's 3D image-taking waypoints and its visiting order for the GTs. Numerical results show that the proposed algorithm significantly reduces the UAV's traveling distance as compared to various benchmark schemes, while meeting the image resolution requirement.
\end{abstract}
\begin{IEEEkeywords}
Unmanned aerial vehicle, traveling salesman problem with neighbourhoods, oblique image acquisition.
\end{IEEEkeywords}

\section{Introduction}
Unmanned aerial vehicles (UAVs) are expected to be widely deployed in future networks to enable various new image acquisition applications such as live broadcast, virtual reality, and so on, by leveraging their advantages of low cost, high mobility, as well as flexible deployment \cite{Zeng1}. To maximize the efficiency of UAV-assisted image acquisition, it is of paramount importance to well design the UAV's three-dimensional (3D) trajectory such that the images of ground targets (GTs) can be captured with minimum traveling distance, while guaranteeing satisfactory image quality. However, unlike the widely studied UAV-enabled communications, this problem still remains unaddressed in the literature, to the authors' best knowledge.

Particularly, there are two challenging issues that need to be resolved in designing the 3D trajectory for UAV-assisted image acquisition. First, prior studies on UAV-assisted image acquisition mostly consider the vertical photography (VP) model, where the UAV-mounted camera is assumed to have a fixed shooting angle which is perpendicular to the ground. As such, the UAV needs to fly above each GT for image acquisition to ensure that the GT is displayed at the center of the captured image \cite{Sai}. However, this strategy may result in long traveling distance and high energy consumption, especially for image acquisition of multiple GTs that are far apart from each other \cite{Mavrinac}. Fortunately, this issue has been resolved in the latest UAV-mounted camera which is able to flexibly adjust its oblique shooting angle according to the GT's location \cite{Hohle}. However, to the best of our knowledge, there still lacks a tractable model to characterize the quality of the images captured by the angle-rotatable camera. Moreover, the trajectory design for UAV-assisted image acquisition is substantially different from that in other missions such as UAV-assisted data collection. Specifically, the UAV should approach each GT such that the captured image can satisfy the resolution requirement, for which the feasible region is generally a complicated function of the GT's location. This makes the trajectory design for taking images of multiple GTs fundamentally different from that for collecting data from multiple ground users (e.g.,  \cite{Zeng2, Lyu}), where the feasible region for the UAV to meet the communication requirement is generally a cylindrical shape with the ground plane centered at the user.

Motivated by the above, we propose in this paper a novel \emph{oblique photography (OP)} model to characterize the resolution of the captured image. Based on this model, we study the 3D UAV trajectory optimization problem to minimize its traveling distance for taking images of multiple GTs, while guaranteeing a minimum resolution requirement of each  captured image. The formulated problem is shown to be a variant of the traveling salesman problem with neighbourhoods (TSPN) \cite{Dumitrscu}, where the neighbourhood represents the feasible region for the UAV to capture the image with satisfactory resolution. Note that although TSPN has been studied in \cite{Zeng2, Lyu} for the two-dimensional (2D) case under disk-shaped neighbourhood, or \cite{Yuan, Isler1} for the 3D case under other regular-shaped neighbourhood, these algorithms cannot be applied to our problem with an irregular neighbourhood region. To tackle this challenging problem, we simplify the UAV trajectory as line segments connected by multiple waypoints, each corresponding to the image-taking location of one GT. Then, we propose an alternating optimization algorithm for finding a suboptimal solution to this simplified problem, by alternately optimizing the waypoint locations and the GT visiting order. Numerical results show that the proposed scheme outperforms various benchmark schemes in terms of the traveling distance, while meeting the image resolution requirement.

\section{System Model and Problem Formulation}
We consider a UAV-assisted image acquisition system with one UAV being dispatched to take images of $K$ GTs, denoted by the set $\mathcal{K}{\rm{=}}\{1,...,K\}$. In the following, we first propose a novel UAV-assisted OP model which is tailored for the angle-rotatable camera, and then formulate the 3D UAV trajectory optimization problem. 
\begin{figure}[htbp!]
\centering
\includegraphics[width=0.5\textwidth]{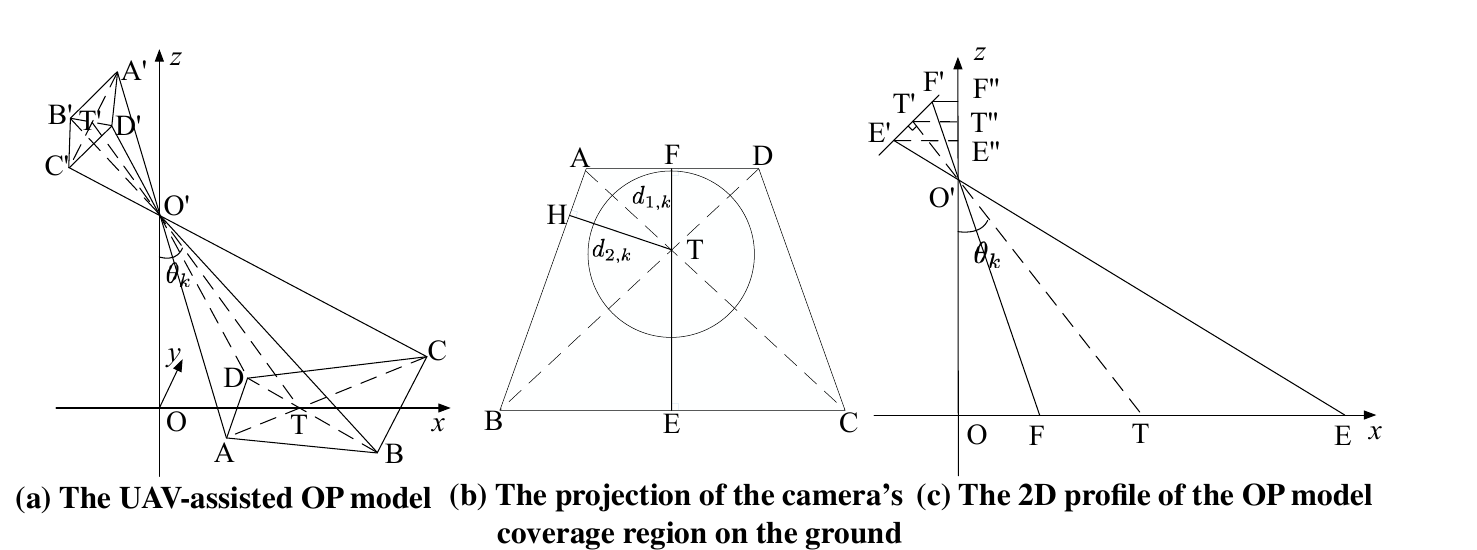}
\caption{Illustration of UAV-assisted OP model.}
\label{F1}
\end{figure}
\subsection{UAV-Assisted OP Model}
In Fig. \ref{F1}(a), we present a UAV-assisted OP model where rectangle { $A'B'C'D'$} is the camera's image plane whose coverage region on the ground is an isosceles trapezoid (i.e., { $ABCD$}). Points { $T'$} and { $T$} represent the centers of the image plane and GT$_k$, respectively. Let $[{{\bf{w}}_k^T,0}]$ in meter (m) denote the 3D coordinate of GT$_{k}$ where ${{\bf{w}}_k}\!\in\!{\mathbb{R}^{2 \times 1}}$ represents its horizontal coordinate. For simplicity, we assume that GT$_{k}$ is a disk with a known radius $r_k$ in\;m. Let ${d_{u, k}}$ denote the distance from the UAV to GT$_{k}$ in\;m, i.e., { $|T'T|$}, which is given by
\begin{equation}\label{E1}
{d_{u, k}} = \sqrt {{{\left\| {{{\textbf{q}}_k} - {{\textbf{w}}_k}} \right\|}^{\rm{2}}}{\rm{ + }}{z_k^2}},
\end{equation}where ${{\textbf{q}}_k}\!\in\!{\mathbb{R}^{2 \times 1}}$ and $z_k$ denote the UAV's horizontal and vertical coordinate when taking image of GT$_k$, respectively.

In the conventional VP model, the image resolution is characterized by ground sample distance (GSD) that each pixel can represent. However, with the angle-rotatable camera, the GSDs that the pixels can represent are different, thus rendering the image resolution representation with GSD inapplicable to the OP model. Therefore, we redefine the image resolution as the ratio of GT$_k$'s area to the camera's coverage area. Specifically, we denote by $f_0$ the camera's focal length, and $w_0$ and $l_0$ the width and length of the image plane, respectively. Let ${\theta_k}$ denote the camera's oblique angle (i.e., { $\angle {OO'T}$} in Fig.  \ref{F1}(a)) and we have $\cos {\theta_k}{\rm{ = }}\frac{{{z_k}}}{{{d_{u, k}}}}$ and $\tan {\theta_k}{\rm{=}}\frac{{\|{{{\textbf{q}}_k} - {{\textbf{w}}_k}}\|}}{{{z_k}}}$.{\footnote{The UAV-mounted camera can rotate its oblique shooting angle towards the GT with the angle $\theta_k$ as shown in Fig. \ref{F1}(a).}} As such, the camera’s coverage area, denoted by $S_k^c$, can be expressed as below with the detailed derivation presented in Appendix {\ref{A}}:  
\begin{equation}\label{E2}
\begin{aligned}
S_k^c = S_k^v\times\phi\left( {{\theta_k}} \right) = \frac{{4z_k^2}}{{b_1b_2}} \times \frac{1}{{{{( {1 - \frac{{1}}{{b_1^2}}{{\tan }^2}{\theta_k}})}^2}{{\cos }^3}{\theta_k}}},
\end{aligned}
\end{equation}
where $b_1{\rm{=}}\frac{2f_0}{w_0}$ and $b_2{\rm{=}}\frac{2f_0}{l_0}$ are constants determined by the camera's parameter setting. Note that $S_k^v$ is the camera's coverage area when taking the image right above GT$_k$, which is proportional to $z_k^2$, and $\phi \left( {{\theta_k}} \right)$ is defined as the {\emph{coverage scaling factor}} which is monotonically increasing with respect to (w.r.t.) ${\theta_k}$. It is worth mentioning that the proposed OP model reduces to the conventional VP model when ${\theta_k} = 0$. Based on the above, the image resolution, denoted by ${{\cal I}_k}$, can be characterized by the UAV's 3D image-taking location as
\begin{equation}\label{E4}
{{\cal I}_k} = \frac{S_{GT_k}}{S_k^c} = \frac{{a_k{{( {z_k^2 -\frac{1}{{{b_1^2}}}{{{\left\| {{{\textbf{q}}_k} - {{\textbf{w}}_k}} \right\|}^{\rm{2}}}}})}^2}}}{{{{( {{{\left\| {{{\textbf{q}}_k} - {{\textbf{w}}_k}} \right\|}^{\rm{2}}}{\rm{ + }}z_k^2})}^{\frac{{\rm{3}}}{{\rm{2}}}}}z_k^{\rm{3}}}},
\end{equation}
where $a_k{=}\frac{{b_1b_2\pi r_k^2}}{{4}}$. In this paper, we consider a minimum resolution requirement for each GT$_k$ denoted by ${\overline{\cal{I}}}_k$, thus the UAV's image-taking location $[{\bf{q}}_k^T, z_k]$ should satisfy ${\cal{I}}_k{\geq} \overline{{\cal{I}}}_k$. Moreover, to let each GT$_k$ be completely projected in the camera's coverage region, $[{\bf{q}}_k^T, z_k]$ should also satisfy $r_k{\leq}\min(d_{1,k},d_{2,k})$, where $d_{1,k}$ and $d_{2,k}$ represent the distances from point $T$ to $AD$ and $BC$ (see Fig.  \ref{F1}(b)), which are defined in Appendix B. It is also worth noting that $[{\bf{q}}_k^T, z_k]$  should satisfy $b_1z_k{-}\|{\bf{q}}_k{-}{\bf{w}}_k\|{\geq}0$ to meet the focal length requirement of the camera, as explained in Appendix A.

\vspace{-5mm}
\subsection{Problem Formulation}
We aim to optimize the 3D UAV trajectory for capturing the images of the $K$ GTs, to minimize the UAV's traveling distance from given initial to final points denoted by $[{\bf{w}}_I^T,z_I]$ and $[{\bf{w}}_F^T,z_F]$, respectively, while ensuring a sufficiently high image resolution for all GTs. Note that since the UAV trajectory is continuous, this problem involves an infinite number of variables, thus making the problem difficult to solve. To simplify the trajectory design, we assume that the UAV trajectory consists of $K{\rm{+1}}$ consecutive line segments in a similar manner as \cite{Zhang}, with $K$ \emph{waypoints} each denoting the UAV's image-taking location for one GT. 

Let $\psi (k)\in\mathcal{K}$ denote the index of the $k$-th visited GT and $[{{\bf{q}}_{\psi (k)}^T},{z_{\psi (k)}}]$ denote the location of the waypoint at which the UAV takes the image of GT$_{\psi(k)}$. For consistence, we define $[{{\bf{q}}_{\psi (0)}^T},{z_{\psi (0)}}] = [{{\bf{w}}_I^T},{z_I}]$ and $[{{\bf{q}}_{{\psi (K+1)}}^T},{z_{{\psi (K+1)}}}]=[{{\bf{w}}_F^T},{z_F}]$. For notational convenience, we define ${\bf{\Psi}}\overset{\Delta}{=}[\psi(1),{...},\psi(K)]$, ${\bf{Q}}\overset{\Delta}{=}[{\bf{q}}_{\psi(1)}^T,...,{\bf{q}}_{\psi(K)}^T]$, and ${\bf{Z}}\overset{\Delta}{=}[z_{\psi(1)},...,z_{\psi(K)}]$. The UAV's traveling distance is thus given by
\begin{equation}\label{E6}
\!\!\!\!\!\!D{({\bf{Q}},{\bf{Z}},{\bf{\Psi}})}{\rm{=}}\!\!\sum\limits_{k = 0}^K\!\!{\sqrt{{{\|{{\bf{q}}_{\psi (k + 1)}}{-}{{\bf{q}}_{\psi (k)}}\|}}^2{+}{{({z_{\psi (k{+}1)}}{-}{z_{\psi (k)}})}}^2}}.
\end{equation}
Let ${\overline{{\cal{I}}}}_{\psi(k)}$ denote the resolution requirement of GT$_{\psi(k)}$. Then, under the given resolution constraints, the 3D UAV trajectory optimization problem can be formulated as
\begin{subequations}
\begin{align}
({\rm{P1}})\mathop {\min }\limits_{\scriptstyle {{{\bf{Q}}}},{\bf{Z}},{\bf{\Psi}} \hfill\atop
\scriptstyle}\;\;\;\;&D{({\bf{Q}},{\bf{Z}},{\bf{\Psi}})}\notag \\
{\rm{s}}{\rm{.t}}{\rm{.}}\;\;\;\;\;&\psi (k) \in {\cal K},~~~~~~~~~~~~&\forall k \in {\cal K},\label{E9a}\\
&\mathop  \cup \limits_{k = 1}^K \psi (k) = {\cal K},&\label{E9b}\\
&{{\cal I}_{\psi (k)}} \ge {\overline {\cal I} _{\psi (k)}},&\forall k \in {\cal K},\label{E9c}\\
&{r_{\psi(k)}} \le \min ( {d_{1,{\psi (k)}},d_{2, {\psi (k)}}}),\!\!&\forall k \in {\cal K},\label{E9d}\\
&b_1{z_{{{\psi (k)}}}}{\rm{-}}\|{{\bf{q}}_{{{\psi (k)}}}}{\rm{-}}{{\bf{w}}_{\psi (k)}}\| {\rm{\geq}} 0,&\forall k\in{\cal K},\label{E9e}
\end{align}
\end{subequations}
where (\ref{E9a})-(\ref{E9b}) specify the feasible set of the GT visiting order ${\bf{\Psi}}$, and (\ref{E9c})-(\ref{E9e}) specify the the feasible region of the UAV's image-taking locations for each GT, which is termed as the ``neighbourhood'' for simplicity. In Fig. {\ref{F2}}, we illustrate the neighbourhood of the GT located at the original point with a minimum resolution requirement of ${\overline {\cal I}} = 0.4$, where the 3D view of the neighbourhood is depicted in Fig. {\ref{F2}}(a), which appears to be a {\emph{spherical sector}}, and the vertical profile of the neighbourhood is shown in Fig. {\ref{F2}}(b), which has the shape of a {\emph{crescent moon}}. Note that unlike UAV-assisted data collection where the communication quality generally increases as the UAV approaches the ground user, the UAV needs to maintain a certain distance from the GT for ensuring the image quality due to the non-convexity of the neighbourhood region as shown in Fig. {\ref{F2}}. It is also observed from Fig. {\ref{F2}}(b) that the image resolution gradually increases from the outside to the inside of the neighbourhood, which intuitively indicates that the UAV may prefer to take the image at the surface of the neighbourhood for reducing its traveling distance. 
\begin{figure}[htbp!]
\centering
\includegraphics[width=0.5\textwidth]{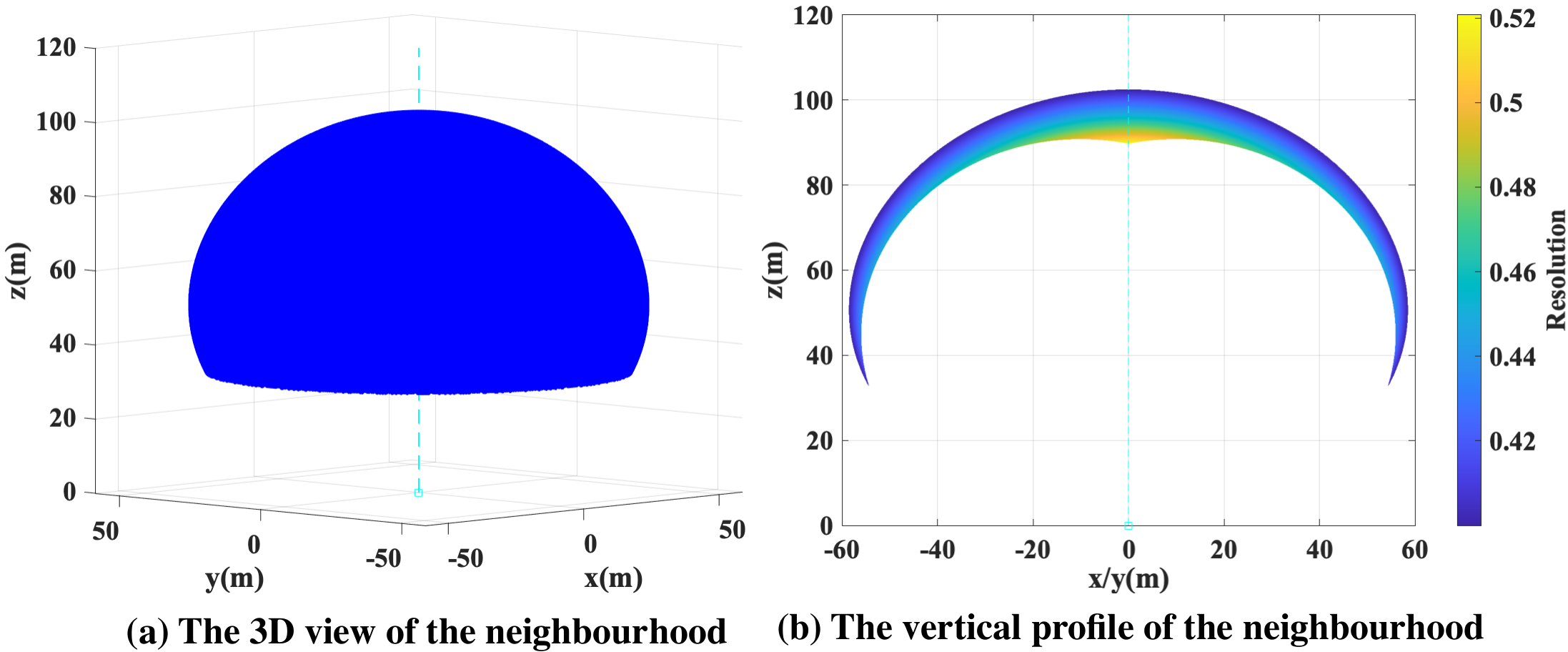}
\caption{Demonstration of the neighbourhood.}
\label{F2}
\end{figure}

Notice that (P1) can be shown to be a modified 3D TSPN problem {\cite{Dumitrscu}}, where the neighbourhood of each GT is a complicated function of the UAV's 3D location as well as the image resolution requirement. It is worth noting that such a 3D TSPN problem is generally NP-hard, and is more involved as compared to the 2D TSPN problem studied in e.g., {\cite{Zeng2}}. In the following, we propose an efficient iterative algorithm for finding a high-quality suboptimal solution to (P1).

\section{Proposed Solution to (P1)}
In this section, we propose an iterative optimization algorithm for solving (P1) by alternately optimizing one between the 3D UAV waypoints and the GT visiting order with the other set of variables being fixed at each time. Specifically, we first introduce the two subproblems, and then present the overall algorithm and analyze its computation complexity. 

\subsection{3D Waypoint Location Optimization}
First, we present the subproblem for optimizing the 3D waypoint locations with given ${\bf{\Psi}}$, which is formulated as
\begin{equation}
\begin{aligned}
({\rm{P2.1}})\;\mathop {\min }\limits_{{\bf{Q}}, {\bf{Z}}} \;\;\;\;&D{({\bf{Q}},{\bf{Z}},{\bf{\Psi}})}\notag\\
{\rm{s}}{\rm{.t}}{\rm{.}}\;\;\;\;&(\ref{E9c}){\rm{-}}(\ref{E9e}). \notag
\end{aligned}
\end{equation}

Since ${\bf{Q}}$ and ${\bf{Z}}$ are coupled with each other and the constraints (\ref{E9c})-(\ref{E9d}) are non-convex, (P2.1) is a non-convex optimization problem, which is difficult to solve in general. To make the problem more tractable, we make some transformations to (\ref{E9c})-(\ref{E9d}). For simplicity, we define ${\boldsymbol{\ell}}_{\psi(k)}\overset{\Delta}{=}{\bf{q}}_{\psi(k)}-{\bf{w}}_{\psi(k)}, \forall k\in \mathcal{K}$, and ${\bf{L}}\overset{\Delta}{=}[{\boldsymbol{\ell}}_{\psi(1)}^T,...,{\boldsymbol{\ell}}_{\psi(K)}^T]$. By taking logarithm of both sides of (\ref{E9c}), we obtain
\begin{equation}\label{E10}
\!\!\!\!\!{\rm-}\frac{3}{2}\ln({{\|{\boldsymbol{\ell}}_{\psi(k)}\|^2}\!{\rm{+}}z_{\psi (k)}^2})\!{\rm{-}}\!3\!\ln {z_{\psi (k)}}\!{\rm{\ge}}f_1(z_{\psi(k)},\!{\boldsymbol{\ell}}_{\psi(k)},\!r_{\psi(k)}),
\end{equation}where
\begin{equation}
\begin{array}{l}
f_1(z_{\psi(k)}, {\boldsymbol{\ell}}_{\psi(k)}, r_{\psi(k)}) \triangleq \ln {\frac{\overline {\cal I} _{\psi (k)}}{a_{\psi(k)}}} - 2\ln( {z_{\psi (k)}^2 {\rm-} \frac{1}{{b_1^2}}{\|{\boldsymbol{\ell}}_{\psi(k)}\|^2}}) \notag
\end{array}
\end{equation}
is a convex function w.r.t. $z_{\psi (k)}$ and $\|{\boldsymbol{\ell}}_{\psi(k)}\|$, respectively. Next, by taking the square of both sides of (\ref{E9d}), (\ref{E9d}) can be equivalently transformed into 
\begin{equation}\label{E11}
\begin{aligned}
\!\!\!z_{\psi (k)}^4{\rm{+}} 2z_{\psi (k)}^2{\|{\boldsymbol{\ell}}_{\psi(k)}\|^2}{\rm{+}} {\|{\boldsymbol{\ell}}_{\psi(k)}\|^4} {\rm{\ge}} f_2(z_{\psi(k)}, {\boldsymbol{\ell}}_{\psi(k)}, r_{\psi(k)}),
\end{aligned}
\end{equation}
where 
\begin{equation}
\begin{aligned}
&f_2(z_{\psi(k)}, {\boldsymbol{\ell}}_{\psi(k)}, r_{\psi(k)}) \triangleq\\
&r_{\psi (k)}^2\max({{{( {b_1{z_{\psi (k)}}{\rm{+}}{\|{\boldsymbol{\ell}}_{\psi(k)}\|}})}^2},
b_2^2z_{\psi (k)}^2{\rm{+}} (1{\rm{+}}b_2^2){{\|{\boldsymbol{\ell}}_{\psi(k)}\|^2}}}) \notag
\end{aligned}
\end{equation}
is a convex function w.r.t. $z_{\psi (k)}$ and $\|{\boldsymbol{\ell}}_{\psi(k)}\|$, respectively. In the following, we apply the block coordinate descent (BCD) technique to decouple the joint optimization for ${\bf{Q}}$ and ${\bf{Z}}$ into two subproblems for ${\bf{Q}}$ and ${\bf{Z}}$, separately, each of which is sub-optimally solved by using the convex approximation technique as in {\cite{You1}}. 

\subsubsection{Optimizing ${\bf{Z}}$ with given ${\bf{Q}}$}
With given ${\bf{Q}}$ and hence ${\bf{L}}$, (P2.1) reduces to the following optimization problem over the altitudes of the $K$ waypoints in ${\bf{Z}}$:
\begin{align}
({\rm{P2.1.1}})\;\mathop {\min }\limits_{{\bf{Z}}} \;\;\;\;&D{({\bf{Q}},{\bf{Z}},{\bf{\Psi}})}\notag\\
{\rm{s}}{\rm{.t}}{\rm{.}}\;\;\;&(\ref{E9e}), (\ref{E10}),(\ref{E11})\notag.
\end{align}

Problem (P2.1.1) is still hard to solve due to the non-convex constraints (\ref{E10})-(\ref{E11}). However, note that with given ${\bf{L}}$, the first term of the left-hand side (LHS) of (\ref{E10}) is convex w.r.t. $z_{\psi (k)}^2$, and so is the second term w.r.t. ${z_{\psi (k)}}$. As such, we can apply the convex approximation technique to approximate the two terms by their lower bounds as follows by using the first-order Taylor expansion at the given local point $z_{\psi (k)}^{(i)}$ of the $i$-th iteration:
{\begin{small}
\begin{align}
&\!\!\!\!\!\!-\!\!\frac{3}{2}\ln ( {{\|{\boldsymbol{\ell}}_{\psi(k)}\|^2}{\rm{ + }}z_{\psi (k)}^2})\!\!\ge\!\! {\varphi _1}({z_{\psi (k)}})\!\! \triangleq \!\!-\!\frac{3}{2}\!\!\ln ( {{\|{\boldsymbol{\ell}}_{\psi(k)}\|^2}{\rm{ + (}}z_{\psi (k)}^{(i)}{)^2}}) \notag \\ 
&\!\!\!\!\!\! -\!\!\frac{3}{{2( {{\|{\boldsymbol{\ell}}_{\psi(k)}\|^2}{\rm{ + (}}z_{\psi (k)}^{(i)}{)^2}})}}({z_{\psi (k)}^2{\rm{-}} {{{\rm{(}}z_{\psi (k)}^{(i)})}^2}}),\label{E12}\\
&\!\!\!\!\!\!-3\ln {z_{\psi (k)}}{\rm{\ge}}{\varphi _2}({z_{\psi (k)}}){\rm{\triangleq}}{\rm{-}}3\ln z_{\psi (k)}^{(i)}{\rm{-}}\frac{3}{{z_{\psi (k)}^{(i)}}}({z_{\psi (k)}}{\rm{-}}z_{\psi (k)}^{(i)}),\label{E13}
\end{align}
\end{small}}where the equality holds at the point $z_{\psi (k)}=z_{\psi (k)}^{(i)}$. With (\ref{E12}) and (\ref{E13}), we approximate (\ref{E10}) with the following constraint:
\begin{equation}\label{E14}
\begin{array}{l}
{\varphi _1}({z_{\psi (k)}}) + {\varphi _2}({z_{\psi (k)}}) \ge f_1(z_{\psi(k)}, {\boldsymbol{\ell}}_{\psi(k)}, r_{\psi(k)}).
\end{array}
\end{equation}
For the constraint (\ref{E11}), it can be shown that its first and second terms on the LHS are both convex w.r.t. ${z_{\psi (k)}}$. This allows us to lower-bound the two terms as follows:
\begin{align}
&\!\!\!\!z_{\psi (k)}^4 \ge {\varphi _3}({z_{\psi (k)}})\!\triangleq\!{{\rm{(}}z_{\psi (k)}^{(i)}{\rm{)}}^4}{\rm{ + 4(}}z_{\psi (k)}^{(i)}{{\rm{)}}^3}({z_{\psi (k)}}{\rm{-}}z_{\psi (k)}^{(i)}), \label{E15}\\
&\!\!2z_{\psi (k)}^2{\|{\boldsymbol{\ell}}_{\psi(k)}\|^2} \ge {\varphi _4}({z_{\psi (k)}}) \notag \\
&\!\!\triangleq 2{{\rm{(}}z_{\psi (k)}^{(i)}{\rm{)}}^2}{\|{\boldsymbol{\ell}}_{\psi(k)}\|^2}{\rm{+}}4z_{\psi (k)}^{(i)}{\|{\boldsymbol{\ell}}_{\psi(k)}\|^2}({z_{\psi (k)}}{\rm{-}}z_{\psi (k)}^{(i)}), \label{E16}
\end{align}
where the equality holds at the point $z_{\psi (k)} = z_{\psi (k)}^{(i)}$. Therefore, we approximate (\ref{E11}) by replacing its LHS with its lower bound as the following constraint:
\begin{equation}\label{E17}
\begin{small}
\begin{array}{l}
\!\!\!\!\!\!\!\!\!\!{\varphi _3}({z_{\psi (k)}}){\rm{+}}{\varphi _4}({z_{\psi (k)}}){\rm{+}}{\|{\boldsymbol{\ell}}_{\psi(k)}\|^4}\!\!\ge\!\!f_2(z_{\psi(k)}, {\boldsymbol{\ell}}_{\psi(k)}, r_{\psi(k)}).
\end{array}
\end{small}
\end{equation}
As such, (P2.1.1) can be reformulated into an approximate form given below, with the LHSs of (\ref{E10}) and (\ref{E11}) replaced by their respective lower bounds:
\begin{align}
({\rm{P2.1.2}})\;\mathop {\min }\limits_{\bf{Z}} \;\;\;\;&D{({\bf{Q}},{\bf{Z}},{\bf{\Psi}})} \notag \\
{\rm{s}}{\rm{.t}}{\rm{.}}~~\;&(\ref{E9e}), (\ref{E14}),(\ref{E17}).\notag \;
\end{align}

(P2.1.2) is a convex optimization problem, which can be efficiently solved via existing software, e.g., CVX. Moreover, it can be shown that the optimal solution to (P2.1.2) is guaranteed to be a feasible solution for (P2.1.1).

\subsubsection{Optimizing ${\bf{Q}}$ with given ${\bf{Z}}$}
With given ${\bf{Z}}$, (P2.1) reduces to the following problem for optimizing the horizontal waypoint locations ${\bf{Q}}$ (or equivalently ${\bf{L}}$):
\begin{equation}
\begin{aligned}
({\rm{P2.1.3}})\;\mathop {\min }\limits_{{{\bf{L}}}} \;\;\;\;&D{({\bf{Q}},{\bf{Z}},{\bf{\Psi}})} \notag \\
{\rm{s}}{\rm{.t}}{\rm{.}}~~\;&(\ref{E9e}),(\ref{E10}),(\ref{E11}).\notag\;\;
\end{aligned}
\end{equation}

Since the first term on the LHS of (\ref{E10}) is convex w.r.t. $\|{\boldsymbol{\ell}}_{\psi(k)}\|^2$, it is lower-bounded by the first-order Taylor expansion at the given local point ${\boldsymbol{\ell}}_{\psi (k)}^{(i)}$ of the $i$-th iteration as
\begin{equation}\label{E18}
\begin{small}
\begin{aligned}
\hspace{-2mm}
&\!\!\!\!\!\!\!\!-\!\frac{3}{2}\ln( {{\|{\boldsymbol{\ell}}_{\psi(k)}\|^2}{\rm{ + }}z_{\psi (k)}^2})\!\ge\!{\vartheta_1}({{{{\boldsymbol{\ell}}}}_{\psi (k)}})\!\!\triangleq\!\!-\!\frac{3}{2}\ln({{\|{\boldsymbol{\ell}}_{\psi(k)}^{(i)}\|^2}{\rm{ + }}z_{\psi (k)}^2})\\
 &\!\!\!\!-\!\frac{3}{{2( {{\|{\boldsymbol{\ell}}_{\psi(k)}^{(i)}\|}^2{\rm{ + }}z_{\psi (k)}^2})}}( {{\|{\boldsymbol{\ell}}_{\psi(k)}\|^2} - {\|{\boldsymbol{\ell}}_{\psi(k)}^{(i)}\|}^2}),
\end{aligned}
\end{small}
\end{equation}where the equality holds at the point ${\boldsymbol{\ell}}_{\psi (k)}{\rm{=}}{\boldsymbol{\ell}}_{\psi (k)}^{(i)}$. Then, (\ref{E10}) can be approximated as
\begin{equation}\label{E19}
\begin{array}{l}
{\vartheta _1}({{{{\boldsymbol{\ell}}}}_{\psi (k)}}) - 3\ln {z_{\psi (k)}} \ge f_1(z_{\psi(k)}, {\boldsymbol{\ell}}_{\psi(k)}, r_{\psi(k)}).
\end{array}
\end{equation}

Similarly, we can derive the lower bounds of the second and third terms on the LHS of (\ref{E11}) as follows by using the first-order Taylor expansion:
\begin{align}
&2z_{\psi (k)}^2{\|\boldsymbol{\ell}_{\psi(k)}\|^2}\ge {\vartheta _2}({{\boldsymbol{\ell}}_{\psi (k)}}) \notag \\
&\triangleq 2z_{\psi (k)}^2{\|{\boldsymbol{\ell}}_{\psi(k)}^{(i)}\|^2}{\rm{+}}4z_{\psi (k)}^2{({{\boldsymbol{\ell}}_{\psi(k)}^{(i)}})^T}({\boldsymbol{\ell}}_{\psi(k)}-{\boldsymbol{\ell}}_{\psi(k)}^{(i)}),\label{E20} \\
&{\|{\boldsymbol{\ell}}_{\psi(k)}\|^4}\ge{\vartheta _3}({{\boldsymbol{\ell}}_{\psi (k)}}) \notag \\
&\triangleq {\|{\boldsymbol{\ell}}_{\psi(k)}^{(i)}\|^4}{\rm{+ }}4{\|{\boldsymbol{\ell}}_{\psi(k)}^{(i)}\|^2}{({{\boldsymbol{\ell}}_{\psi(k)}^{(i)}})^T}({\boldsymbol{\ell}}_{\psi(k)}-{\boldsymbol{\ell}}_{\psi(k)}^{(i)}). \label{E21}
\end{align}
With (\ref{E20}) and (\ref{E21}), the constraint in ({\ref{E11}}) is approximated as follows by replacing its LHS with its lower bound:
\begin{equation}\label{E22}
\begin{array}{l}
\!\!\!\!z_{\psi (k)}^4{\rm{ + }}{\vartheta _2}({{\boldsymbol{\ell}}_{\psi (k)}}){\rm{+}} {\vartheta _3}({{\boldsymbol{\ell}}_{\psi (k)}})\!\ge\!f_2(z_{\psi(k)}, {\boldsymbol{\ell}}_{\psi(k)}, r_{\psi(k)}).
\end{array}
\end{equation}
As a result, (P2.1.3) can be reformulated into the following approximate form:
\begin{equation}
\begin{aligned}
({\rm{P2.1.4}})\;\mathop {\min }\limits_{{{\bf{L}}}} \;\;\;\;&D{({\bf{Q}},{\bf{Z}},{\bf{\Psi}})} \notag \\
{\rm{s}}{\rm{.t}}{\rm{.}}\;\;\;\;&(\ref{E9e}), (\ref{E19}),(\ref{E22}).\notag\;\;\;
\end{aligned}
\end{equation}

Since the constraints in (\ref{E19}) and (\ref{E22}) are convex, (P2.1.4) is a convex optimization problem, which can be efficiently solved via existing software, e.g., CVX. 

With the convex approximation technique, the objective value of (P2.1) can be shown to be non-increasing over the iterations similarly as in \cite{You1}, which is also lower-bounded by a finite value. Therefore, the proposed algorithm for optimizing the 3D waypoint locations is guaranteed to converge.

\subsection{Visiting Order Optimization}
With given waypoint locations $({\bf{Q}},{\bf{Z}})$, the subproblem for optimizing the GT visiting order ${\bf{\Psi}}$ is recast as
\begin{align}
({\rm{P2.2}})\;\mathop {\min }\limits_{{\bf{\Psi}}} \;\;\;\;&D{({\bf{Q}},{\bf{Z}},{\bf{\Psi}})} \notag \\
\;\;\;\;\;\;\;{\rm{s}}{\rm{.t}}{\rm{.}}\;\;\;\;&(\ref{E9a}){\rm{,}}(\ref{E9b}) \notag.
\end{align}

(P2.2) is equivalent to a classic TSP, for which a high-quality suboptimal solution can be found with low computational complexity via binary integer programming \cite{Miller}.

\subsection{Overall Algorithm and Computational Complexity}
\begin{figure*}[htbp!]
\centering
\includegraphics[width=0.65\textwidth]{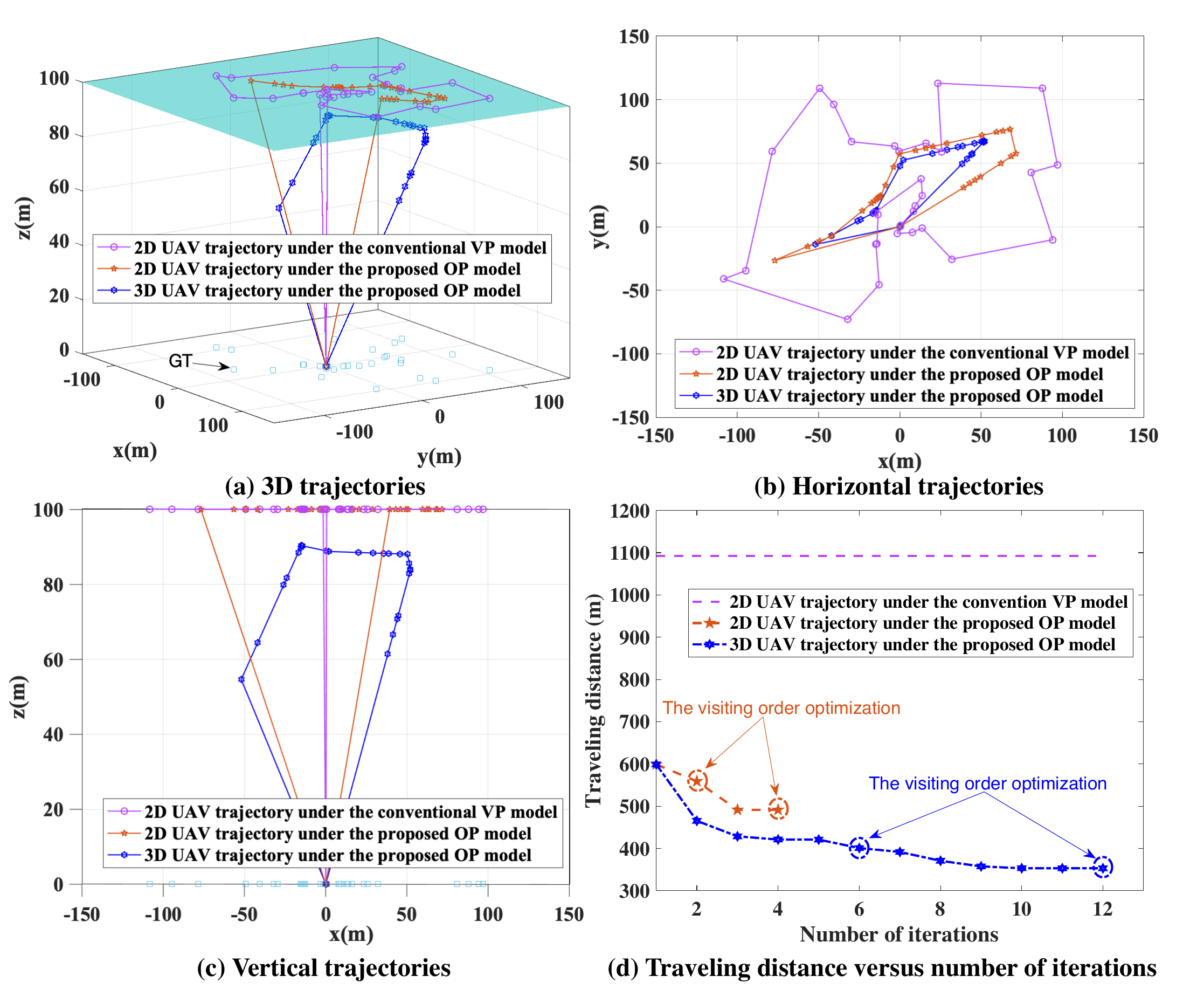}
\caption{Comparison of the optimized UAV trajectories and traveling distances by different schemes.}
\label{F3}
\end{figure*}
The proposed algorithm for (P1) is summarized as follows.{\footnote{The proposed algorithm can be generally extended to solve the TSPNs in e.g.,  \cite{Lyu, Yuan, Isler1} by modifying the neighbourhood-related constraints in (P1).}} First, we initialize the visiting order $\bf{\Psi}$ by solving the TSP in (P2.2) based on ${\bf{q}}_k={\bf{w}}_k, z_k=0, \forall k\in \mathcal{K}$, i.e., considering waypoint locations at the GTs. Then, we iteratively optimize the 3D waypoint locations and GT visiting order based on Sections III-A and III-B, respectively. Next, with given $\bf{Q}$ and $\bf{Z}$, $\bf{\Psi}$ is optimized by solving (P2.2). The proposed algorithm stops until we cannot find a better solution within a prescribed precision requirement or a maximum number of iterations is reached. The overall complexity of the proposed algorithm is analyzed as follows. For the subproblem of waypoint locations optimization, ${\bf{Q}}$ and ${\bf{Z}}$ are iteratively optimized by using the convex software based on the interior-point method, and their individual complexity can be represented as $\mathcal{O}(K^{3.5})$ and $\mathcal{O}(K^{3.5})$, respectively. Then, let $I_1$ denote the number of iterations for the BCD method, the total computation complexity for optimizing the waypoint locations is $\mathcal{O}(K^{3.5}I_1)$. For the subproblem of visiting order optimization, the complexity for solving the classic TSP with the algorithm in \cite{Miller} is $\mathcal{O}(2^KK^2)$. As such, the overall complexity is $\mathcal{O}\left((2^KK^2{\rm{+}}K^{3.5}I_1)I_2\right)$, where $I_2$ denotes the number of inter-subproblem iterations.

\section{Numerical Results}
In this section, we provide numerical results to show the effectiveness of the proposed OP model as well as the corresponding 3D trajectory design. The parameters are set as $f_0{=}0.035$\;{m}, $w_0{=}0.0156$\;m, $l_0{=}0.0235$\;m {\cite{Sun}}, and $[{{\bf{w}}_I^T},{z_I}]{\rm{=}}[{{\bf{w}}_F^T},{z_F}]{\rm{=}}[0, 0, 0]$\;m. We consider $30$ GTs with the same radius of $r_k{\rm{=}}20$\;m, $k\!\in\!\mathcal{K}$, which are randomly distributed in a square area of $300{\rm{\times}}300$\;m$^2$. The resolution requirement of each GT, i.e., $\overline{\mathcal{I}}_k$, $k\!\in\!\mathcal{K}$, is independently and randomly set within $[0.01,0.4]$. Two benchmark schemes are considered: 1) 2D UAV trajectory under the conventional VP model, and 2) 2D UAV trajectory under the proposed OP model. For the two benchmark schemes, the UAV altitude is fixed (i.e., $z_k = 100$\;m, $\forall k \in {\cal K}$) to ensure that at least a feasible waypoint can be found for each GT to satisfy the image resolution requirement. Specifically, in the benchmark scheme 1, the image-taking waypoint for each GT is right above the GT and thus the UAV trajectory can be obtained by solving the TSP problem based on these waypoints. While in the benchmark scheme 2, the UAV trajectory can be obtained via the iterative algorithm proposed in Section III by alternately optimizing the horizontal waypoint locations in ${\bf{Q}}$ and the GT visiting order in ${\bf{\Psi}}$.  

Figs. \ref{F3}(a)-(c) show the UAV trajectories obtained by three schemes. It is observed from Fig. \ref{F3}(b) that compared to the UAV trajectory under the conventional VP model, the UAV's horizontal flight range can be greatly reduced by adopting the proposed OP model, even with the fixed altitude (as in benchmark scheme 2). Moreover, it is observed from Fig. \ref{F3}(c) that under the proposed OP model, the 3D UAV trajectory in general has a lower altitude than the 2D UAV trajectory. This is because the UAV can also satisfy the resolution requirement when taking the image of the GT at a lower altitude by exploiting a larger horizontal distance away from the GT, thus resulting in a shorter traveling distance, which can be inferred from Fig. {\ref{F2}}(b). Fig. \ref{F3}(d) shows the traveling distances versus the number of iterations for all schemes. It is observed that the proposed OP model with 2D trajectory yields much shorter distance than the conventional VP model with 2D trajectory, which is further reduced by the proposed OP model with 3D trajectory due to additional degrees-of-freedom in the vertical trajectory optimization (see Fig. \ref{F3}(c)). Specifically, the circled dots denote the traveling distance obtained by optimizing the GT visiting order, while other dots correspond to the traveling distances obtained by optimizing the horizontal/vertical waypoint locations in ${\bf{Q}}$ or ${\bf{Z}}$.

\section{Conclusions}
In this paper, we proposed a novel OP model to characterize the resolution of images captured by an angle-rotatable camera mounted on a UAV. Under the proposed OP model, we formulated a 3D UAV trajectory optimization problem to minimize the UAV's traveling distance while maintaining a given resolution requirement for the images taken from multiple GTs. The formulated problem was shown to be a modified 3D TSPN problem, for which we proposed an iterative algorithm for finding an efficient solution, by alternately optimizing the image-taking waypoints and the visiting order of the GTs. Numerical results were presented to show the effectiveness of the proposed scheme compared to other benchmark schemes.  

\begin{appendices}
\section{}\label{A}
For ease of explanation, we illustrate in Fig. \ref{F1}(c) the 2D profile of Fig. \ref{F1}(a). Specifically, we have {$|E'F'|{\rm{=}}w_0$}, {$|T'O'|{\rm{=}}f_0$}, and {$\angle {OO'T}{\rm{=}}{\theta_k}$}. According to the {\emph{triangle similarity theorem}} (TST), we have {$\frac{{|{O'{E''}}|}}{{|O'O|}}{\rm{=}}\frac{{|E'E''|}}{{|{{O}E}|}}$} where {$|{{O'{O}}}|{\rm{=}}({{d_{u, k}}{\rm{-}}{f_0}})\cos {\theta_k}$, $|{{O'{E''}}}|{\rm{=}}{f_0}\cos {\theta_k}{\rm{-}} \frac{{{w_0}}}{2}\sin {\theta_k}$}, and {$|{{E'{E''}}}|{\rm{=}}{f_0}\sin {\theta_k}{\rm{+}}\frac{{{w_0}}}{2}\!\cos {\theta_k}$}. To ensure {$|{O'E''}|{\rm{\geq}}0$, ${\theta_k}$} should satisfy that $0{\rm{\leq}}{\theta_k}{\rm{\leq}}\arctan\frac{{2f_0}}{w_0}$, which is equivalent to
\begin{equation}\label{E23}
b_1z_k-{\left\| {{{\textbf{q}}_k}-{{\bf{w}}_k}} \right\|}\geq 0,
\end{equation}
where $b_1=\frac{2f_0}{w_0}$. Then, {${{|OE|}}$} can be obtained as
\begin{equation}\label{E24}
|{{OE}}| = \left( {{d_{u,k}} - {f_0}} \right)\cos {\theta_k} \times \frac{{{f_0}\sin {\theta_k} + \frac{{{w_0}}}{2}\cos {\theta_k}}}{{{f_0}\cos {\theta_k} - \frac{{{w_0}}}{2}\sin {\theta_k}}}.
\end{equation}

Similar to {$|{OE}|$}, { $|{OF}|$} is given by
\begin{equation}\label{E25}
|{{OF}}| = \left( {{d_{u,k}} - {f_0}} \right)\cos {\theta_k} \times \frac{{{f_0}\sin {\theta_k} - \frac{{{w_0}}}{2}\cos {\theta_k}}}{{{f_0}\cos {\theta_k} + \frac{{{w_0}}}{2}\sin {\theta_k}}}.
\end{equation}

Since { $|{EF}|\!=\!|{OE}|\!-\!|{OF}|$}, { $|{EF}|$} can be derived as
\begin{equation}\label{E26}
|{{EF}}| = \frac{{{f_0}{w_0}\left( {{d_{u,k}} - {f_0}} \right)\cos {\theta_k}}}{{f_0^2{{\cos }^2}{\theta_k} - \frac{{w_0^2}}{4}{{\sin }^2}{\theta_k}}}.
\end{equation}

Similarly, { $|{AD}|$} and { $|{BC}|$} can be obtained as follows with the derivation omitted for brevity.
\begin{small}
\setlength{\abovedisplayskip}{1pt}
\setlength{\belowdisplayskip}{1pt}
\begin{align}
\!\!\!\!\!\!|{{BC}}| = \frac{{{l_0}\left( {{d_{u,k}} - {f_0}} \right)\cos {\theta_k}}}{{{f_0}\cos {\theta_k} - \frac{{{w_0}}}{2}\sin {\theta_k}}}, 
|{{AD}}| = \frac{{{l_0}({{d_{u,k}} - {f_0}})\cos {\theta_k}}}{{{f_0}\cos {\theta_k} + \frac{{{w_0}}}{2}\sin {\theta_k}}}. \label{E28}
\end{align}
\end{small}
%\begin{align}
%\!\!\!\!\!\!|{{BC}}| = \frac{{{l_0}\left( {{d_{u,k}} - {f_0}} \right)\cos {\theta_k}}}{{{f_0}\cos {\theta_k} - \frac{{{w_0}}}{2}\sin {\theta_k}}}, 
%%\end{align}
%%\begin{align}
%|{{AD}}| = \frac{{{l_0}({{d_{u,k}} - {f_0}})\cos {\theta_k}}}{{{f_0}\cos {\theta_k} + \frac{{{w_0}}}{2}\sin {\theta_k}}}. \label{E28}
%\end{align}

According to the trapezoid area formula, i.e., { $S_k^c = \frac{1}{2} \cdot |{EF}| \cdot (|{AD}| + |{BC}|)$}, the camera's coverage area is given by
\begin{equation}\label{E29}
S_k^c =  \frac{{4{({d_{u,k}} - {f_0})}^2}}{{b_1b_2}} \times \frac{1}{{{{( {1 - \frac{{1}}{{b_1^2}}{{\tan }^2}{\theta_k}})}^2}{{\cos }}{\theta_k}}},
\end{equation}
where $b_2 = \frac{2f_0}{l_0}$. Since {$f_0\ll{d_{u,k}}$}, we have {${\left( {{d_{u,k}}-{f_0}} \right)^2}\approx d_{u,k}^2 = \frac{{z_k^2}}{{{{\cos }^2}{\theta_k}}}$}. Thus, (\ref{E29}) can be rewritten as
\begin{equation}\label{E30}
S_k^c \approx \frac{{4z_k^2}}{{b_1b_2}} \times \frac{1}{{{{( {1 - \frac{{1}}{{b_1^2}}{{\tan }^2}{\theta_k}})}^2}{{\cos }^3}{\theta_k}}}.
\end{equation}

\section{}\label{B}
As shown in Fig. \ref{F1}(b), it is required that both { $|{FT}|$} (denoted by { $d_{1, k}$}) and { $|{HT}|$} (denoted by { $d_{2, k}$}) derived in the following should be no smaller than $r_k$. According to the TST, we have { $\frac{|{AD}|}{|{BC}|}=\frac{|{FT}|}{|{ET}|}$} where { $|{ET}|+|{FT}|=|{EF}|$}. Therefore, we obtain
\begin{equation}\label{E31}
d_{1, k}=|{FT}|=\frac{|{AD}|\times|{EF}|}{|{AD}|+|{BC}|}.
\end{equation}

With { $|{AD}|$}, { $|{BC}|$}, and { $|{EF}|$} given in Appendix {\ref{A}}, { $d_{1,k}$} can be approximated as
\begin{equation}\label{E32}
d_{1, k} \approx \frac{{z_k^2 + {{\left\| {{{\textbf{q}}_k} - {{\textbf{w}}_k}} \right\|}^{\rm{2}}}}}{{b_1{z_k} + \left\| {{{\textbf{q}}_k} - {{\textbf{w}}_k}} \right\|}}.
\end{equation}

Similar to $d_{1,k}$, $d_{2,k}$ can be denoted by 
\begin{equation}\label{E35}
d_{2, k} {\rm{=}} |{HT}| = \frac{|{AD}|\times|{BC}|\times|{EF}|}{(|{AD}|+|{BC}|)\times|{AB|}},
\end{equation}
where $|{AB}| = \sqrt{{(|{BC}|-|{AD}|)^2}/{4}+|{EF}|^2}$. With { $|{AD}|$}, { $|{BC}|$}, and { $|{EF}|$} given in Appendix {\ref{A}}, $d_{2, k}$ can be approximated as
\begin{equation}\label{E36}
d_{2, k} \approx \frac{{z_k^2 + {{\left\| {{{\textbf{q}}_k} - {{\textbf{w}}_k}} \right\|}^{\rm{2}}}}}{{{{( {b_2^2z_k^2 + (1 + b_2^2){{\left\| {{{\textbf{q}}_k} - {{\textbf{w}}_k}} \right\|}^2}})}^{\frac{1}{2}}}}}.
\end{equation}

Please note that we approximate { ${d_{u,k}}{\rm{-}}{f_0}$ as ${d_{u,k}}$} in (\ref{E32}) and (\ref{E36}) as that in Appendix {\ref{A}} for simplicity.
\end{appendices}

\section*{Acknowledgement}
This work was supported in part by the National Natural Science Foundation of China under Grants 62101474 and 62071332, in part by Shanghai Rising-Star Program under Grant No. 19QA1409100, and in part by the Fundamental Research Funds for the Central Universities.

\vfill

\bibliographystyle{IEEEtran}
\end{document}